\documentclass[12pt]{article}
\begin{document}
\vbadness = 100000
\hbadness = 100000
\title{Experimental tests of sum rules
for heavy baryon masses}
\author{Jerrold Franklin\footnote{Internet address:
Jerry.F@TEMPLE.EDU}\\
Department of Physics\\
Temple University, Philadelphia, PA 19122-6082
\date{November 13, 2008}}
\maketitle

\begin{abstract}

Model independent sum rules for heavy baryon masses are tested for baryons containing  charmed or bottom quarks.  
The sum rules depend only on the assumption that baryon mass differences are determined by spin-dependent two-body energies of quark pairs that do not depend on which baryon the quark pair is in.  No assumption is made about the details of the quark-quark interaction.  The sum rules are generally well satisfied, although better experimental accuracy would be required for a quantitative test of some of the sum rules.
The success of the sum rules is evidence that ``baryon independence" of quark-quark interaction is a good assumption for baryon mass calculations.  It also means that the success of some detailed baryon mass calculations may not depend on the specific mechanism used in the calculation.
\end{abstract}

PACS numbers: 12.40.Yx., 13.40.Dk, 14.20.Lq, 14.20.Mr
\vspace{.5in}

\section{Introduction}

Model independent sum rules[1-5] were derived some time ago for heavy-quark
baryon masses using fairly minimal assumptions within the quark model.
The sum rules depend on standard quark model
assumptions, and an additional assumption that the interaction
energy of a pair of quarks in a particular spin state does not
depend on which baryon the pair of quarks is in (``baryon independence").
This is a somewhat weaker assumption than full SU(3) symmetry of the wave
function, which would require the same spatial wave function for each octet
baryon, and for each individual wave function to be SU(3) symmetrized.
Instead, we use wave functions with no SU(3) symmetry, as described in Ref.\cite{sqm}.
In deriving the sum rules, no assumptions are made about the type of potential,
and no internal symmetry beyond baryon independence is assumed.
The sum rules allow any amount of symmetry breaking in the interactions and
individual wave functions, but do rest on baryon independence for each
quark-quark interaction energy.  Some of the sum rules, Eqs.\ (1, 2, 3, 10, 12),
allow for an orbital component of the three quark wave function.  This is because the q-q interaction energies in these sum rules depend only on the position of the quarks in the baryon wave function.  Sum rules connecting spin $\frac{3}{2}$ baryons with spin $\frac{1}{2}$ baryons require that there be no orbital excitation. 
More detailed discussion of the derivation of the sum rules is given in I.

We have previously tested some of the sum rules in Refs.\ [2-5]
using early measurements of heavy baryon masses.  Those tests showed
reasonable agreement within fairly large experimental errors  for the sum rules tested,
while other sum rules made predictions of heavy baryon masses.  
Since then there have been  new measurements,\cite{pdg} resulting in more accurate and reliable values for a large number of charmed and bottom baryon masses.  
In this paper we test the sum rules in light of these of the new measurements.

\section{Mass differences within isotopic multiplets}
 
We first give isospin breaking sum rules for mass differences within isotopic multiplets.  These are expected to depend on Coulomb, magnetic moment, and QCD spin-spin interactions, but we make no assumption here about the form of these interactions.  From Eq.\ (10) of I, we have
\begin{eqnarray}
d-u+dd-uu&=&\Sigma^+ +\Sigma^-  -2\Sigma^0\quad(1.7\pm 0.2)\nonumber\\
 &=& \Sigma^{*+}+\Sigma^{*-}-2\Sigma^{*0}\quad(2.6\pm 1.2)\nonumber\\
& =& \Sigma_c^{++}+\Sigma_c^0-2\Sigma_c^+\quad(2.1\pm 0.8)\nonumber\\
& =& \Sigma_c^{*++}+\Sigma_c^{*0}-2\Sigma_c^{*+}.\quad(2.5\pm 4.7)
\label{eq:sigmas}
\end{eqnarray}
We have written the experimental values (in MeV) for each sum in parentheses following each equation.
The experimental masses for the heavy baryons are taken from the 2008 PDG summary.\cite{pdg}
The quark energy combination for each sum is given on the left hand side of the equation.  A single quark symbol represents the quark mass, while each quark pair ({\it e.g.}$dd$) represents the quark-quark interaction energy.
The interaction energies depend on whether the q-q spin is zero or one or a mixed spin state.
Unless we indicate otherwise, the q-q combinations on the left hand side of the sum rules are in the spin one state.

The sum rule of Eq.\ (1), and all our charmed baryon sum rules, also apply to bottom baryons with the simple substitution of c$\rightarrow$b in the quark content in any of the sum rules.
Not enough bottom baryon masses have been measured to test the bottom baryon analogues of Eq.\ (1), but they can be used to predict the 
$\Sigma^0_b$ and $\Sigma^{*0}_b$ masses as
\begin{eqnarray}
\Sigma_b^0 &=& \frac{1}{2}(\Sigma_b^+ +\Sigma_b^- -1.7)=5812\pm 2,\\
 \Sigma_b^{*0} &=& \frac{1}{2}(\Sigma_b^{*+} +\Sigma_b^{*-}-1.7)=5833\pm 2.
\end{eqnarray}

Equation (11) of I relates the mass differences of $\Sigma^*$ and $\Xi^*$ baryons as
\begin{eqnarray}
uu-dd+2(ds-us)&=&(\Sigma^{*+}-\Sigma^{*-}) + 2(\Xi^{*-}-\Xi^{*0})\quad(2.0\pm 1.3)\nonumber\\
&=&(\Sigma_c^{*+}-\Sigma_c^{*-}) + 2(\Xi_c^{*-}-\Xi_c^{*0}).\quad(-0.7\pm 1.8)
\end{eqnarray}
The experimental errors on these mass differences
are still too large at this point to make an accurate comparison with
experiment.

Equation (14) of Ref.\ \cite{cb2} was not included in I because it has no purely light baryon counterpart.  
We used it in Ref.\ \cite{cb2} to predict the $(\Xi_c^{'0}-\Xi_c^{'+})$ mass difference, which has since been measured.
The sum rule can be written as
\begin{eqnarray}
(us-ds)+(dc'-uc')&=&(\Xi_c^{'0}-\Xi_c^{'+}) + (\Xi^{*0}-\Xi^{*-})\quad(-0.1\pm 0.8)
\nonumber\\
&=&\frac{1}{2}\left[(\Sigma^{*+}-\Sigma^{*-})+(\Sigma_c^{0}-\Sigma_c^{++})\right].
(-1.8\pm 0.3)
\label{eq:xi}
\end{eqnarray}
The combinations $dc'$ and $uc'$ are the interaction energies of mixed spin states of nucleon and charmed quarks, which are assumed to be the same in the $\Xi'_c$ and $\Sigma_c$ baryons.  Our notation is that the $\Xi'_c$ baryon has the nucleon and strange quarks in the spin one state.  This is the opposite of our usage in I, but was followed in Refs.[2-5], and is now the notation in the PDG tables.

Equation (12) of I involves eight charmed baryons, and
is our only sum rule involving $\Xi_c$, which has its nucleon and strange quarks in the spin zero state.  It reads
\begin{eqnarray}
&11(p-n)+6(\Sigma^--\Sigma^+)+(\Xi^{*0}-\Xi^{*-})\hspace{1.6in}(31\pm 1)&\nonumber\\
&=(\Xi_c^0-\Xi_c^+)+9(\Xi_c^{'0}-\Xi_c^{'+})+2(\Sigma_c^{++}-\Sigma_c^{0})
+3(\Sigma_c^{*++}-\Sigma_c^{*0}).\hspace{.1in}(31\pm 5)&
\end{eqnarray}
  
\section{Strong interaction mass differences}

Equations (4) and (5) of Ref.\ \cite{cb3} can be written as
\begin{eqnarray}
3(ud_1-ud_0)&=&2(\Sigma^{*0}-\Lambda^0)+(\Sigma^0 -\Lambda^0)\quad(614\pm 2)\nonumber\\
&=&2(\Sigma_c^{*+}-\Lambda_c^+)+(\Sigma_c^+ -\Lambda_c^+)\quad(631\pm 1)\nonumber\\
&=&2(\Sigma_b^{*0}-\Lambda_b^0)+(\Sigma_b^0 -\Lambda_b^0).\quad(615\pm 1)
\label{eq:ud}
\end{eqnarray}
We have used Eqs.\ (2) and (3) to find the neutral $\Sigma_b$ baryons from the charged ones.
 The sum rule of Eq.\ (6) in Ref.\cite{cb3} can now be tested with improved precision:
\begin{eqnarray}
uu+ss-2us&=& \Sigma^+ +\Omega^{-} -\Xi^0 -\Xi^{*0}\quad(14)\nonumber\\
&=&\Sigma_c^{++}+\Omega_c^{0} -2\Xi_c^{' +}\quad(1\pm 4)\nonumber\\
&=&\Sigma_c^{*++}+\Omega_c^{*0} -2\Xi_c^{* +}.\quad(-7\pm 3)
\label{eq:us}
\end{eqnarray}
Equation (7) of Ref.\ \cite{cb4} was originally used to predict the $\Omega_c^{*0}$
mass, but now gives the sum rule
\begin{eqnarray}
(\Omega_c^{*0}-\Omega_c^0)+(\Sigma_c^{*++}-\Sigma_c^{++})-2(\Xi_c^{*+}-\Xi_c^{'+})=0.
\quad(-7\pm 4)&&
\end{eqnarray}
The sum rules in Eqs.\ (5-9) are satisfied to about the same
extent as light-quark baryon sum rules relating spin-$\frac{1}{2}$ baryon masses
to spin-$\frac{3}{2}$ baryon masses.\cite{cb,cb3,sqm}

\section{Doubly charmed baryons} 
There also are sum rules that can be used to predict mass differences of doubly charmed (or bottom) baryons. 
Equations (9), (13), and (15) of I can be used to predict the mass differences
\begin{eqnarray}
\Xi_{cc}^{++}-\Xi_{cc}^+ &=& (\Sigma_c^{++}-\Sigma_c^0)+(n-p)=1.6\pm 0.1.\\
\Xi_{cc}^{*++}-\Xi_{cc}^{*+} &=& (\Sigma_c^{*++}-\Sigma_c^{*0})+(n-p)=1.6\pm 0.6\\
\Omega_{cc}^+ -  \Xi_{cc}^{++}&=&(\Omega_c^0-\Sigma_c^{++})-( \Xi^0-\Sigma^+)=118\pm 3\\
\Omega_{cc}^{*+} -  \Xi_{cc}^{*++}&=&(\Omega_c^{*0}-\Sigma_c^{*++})-( \Xi^{*0}-\Sigma^{*+})
=106\pm 3\\
\Omega_{cc}^{*+}-\Omega_{cc}^{+}&=&\Omega_c^{*0}-\Omega_c^{0}=71\pm 2.
\end{eqnarray}
Equations (16) and (17) of I could be used to predict the masses of  triply charmed Omega baryons if enough
doubly charmed baryons are found.  As mentioned previously, all of the above charmed baryon equations would also apply for bottom baryons.

\section{Conclusion}

In conclusion, we can say that increasingly accurate experimental mass
determinations are making the model independent sum rules discussed here increasingly
useful tests of the quark model for heavy baryons.
The baryon mass sum rules are generally satisfied, although increased experimental precision is needed to make significant tests of some of the sums. 
The success of the sum rules gives some confidence that baryon independence, if used in a detailed mass calculation, is a reasonable assumption for that purpose.  This also means that the success of a detailed mass calculation may not mean that the detailed mechanism is correct.

We should point out that sum rules for baryon magnetic moments\cite{mom} are not as successful, having some significant disagreements with experiment.  We take this as an indication that the magnetic moments are a more sensitive test of baryon composition than mass calculations.  While baryon independence seems to work for baryon masses, this does not necessarily imply that it is  a general property of the detailed structure of baryons.

\end{document}